\documentclass[%
 aip,
 sd,%
 amsmath,amssymb,
 preprint
]{revtex4-1}

\usepackage{graphicx}
\usepackage{dcolumn}
\usepackage{bm}

\usepackage{xspace}
\def\locdim{d}
\def\intdim{\mathcal{D}}
\def\comgood{\,,}
\def\dotgood{\,.}

\begin{document}

\title[Local dimension and recurrent circulation patterns in long-term climate simulations.]{Local dimension and recurrent circulation patterns in long-term climate simulations.}

\author{Sebastian Buschow}
 \affiliation{ 
 	Meteorological Institute, University of Bonn, 53121 Bonn, Germany
 }%
\author{Petra Friederichs}%
\affiliation{ 
	Meteorological Institute, University of Bonn, 53121 Bonn, Germany
}%

\date{\today}

\begin{abstract}
With the recent advent of a sound mathematical theory for extreme events in dynamical systems, new ways of analyzing a system's inherent properties have become available: Studying only the probabilities of extremely close Poincar\'{e} recurrences, we can infer the underlying attractor's local dimensionality -- a quantity which is closely linked to the predictability of individual configurations, as well as the information gained from observing them.
This study examines possible ways of estimating local and global attractor dimensions, identifies potential pitfalls and discusses conceivable applications. The Portable University Model of the Atmosphere (PUMA) serves a test subject of intermediate complexity between simple mathematical toys and truly realistic atmospheric data-sets. It is demonstrated that the introduction of a simple, analytical estimator can streamline the procedure and allows for additional tests of the agreement between theoretical expectation and observed data. We furthermore show how the newly gained knowledge about local dimensions can complement classical techniques like principal component analysis and may assist in separating meaningful patterns from mathematical artifacts. 
\end{abstract}


\maketitle

\begin{quotation}
Climate scientists have long been interested in the properties of attracting sets in atmosphere-like dynamical systems. One of the most basic characteristics of interest is the attractor's dimension -- intuitively relating to the number of directions in phase space in which the system may evolve and the amount of information needed to describe individual configurations. It is believed that an attractor's local dimensionality characterizes different dynamical regimes and can be used to asses instantaneous predictability. Using recent results of extreme value theory, this quantity can, in principle, be inferred from observed cases of extremely close recurrence. In this study, we suggest a new way of exploiting these results, which simplifies the procedure and enables us to assess convergence to the theoretically predicted behavior in cases where the true limit dimension is unknown. A climate model of intermediate complexity serves as a convenient test case. It is shown that convergence is not achieved, but experimentally estimated dimensions are nonetheless tied to distinct regimes of circulation. 
\end{quotation}

\section{Introduction}\label{sec:intro}
When attempting to forecast future states of chaotic systems like the earth's atmosphere, the question of predictability is of high interest. Since E.N.\,Lorenz' seminal paper \cite{Lorenz1963}, atmospheric scientists have been well aware that deterministic forecasts are fundamentally limited: While the laws of physics enable us to understand the relevant dynamics and make  accurate predictions for a few days, forecast skill soon vanishes due to uncertainties in model formulation and initial condition. In order to quantify the problem and possibly guide our efforts towards amelioration,  alternative approaches are needed.\par
In this paper, we attempt to exploit results of dynamical system theory, starting with the assumption that the time evolution in our model preserves the natural probability measure. Simply put, we demand that the probability of a set on initial conditions equals the probability of the set of first iterates. The typical observed behavior of atmosphere-like systems is then represented by the following further model assumptions:
Firstly, if two initial conditions are separated by a small vector $\varepsilon_0$, the distance vector after $n$ steps initially evolves exponentially, i.e. $\varepsilon_n=\varepsilon_0\lambda^n$. Here, $\lambda$ is the Lyapunov number for the direction of $\varepsilon_0$. For a system to be chaotic, at least one direction must exhibit exponential error \textit{growth}, i.e. at least one $\lambda$ is greater than zero. In high-dimensional settings, the sum over all positive $\lambda_i$ plays an important role since it provides an upper bound on the system's metric entropy, intuitively characterizing the degree of chaos\cite{eckmann1985}.\par 

Secondly, physics dictates that not all conceivable system configurations are realized and typical initial states don't diverge indefinitely. Instead, most trajectories are drawn towards a subset $A\subset\Omega$, eventually approaching it arbitrarily closely and never leaving its vicinity. The set $A$, called the system's \textit{attractor}, could be a simple geometric object like a point (a so-called fixed point) or a closed loop (limit cycle). In the case of deterministic chaos, $A$ is typically a \textit{fractal} \cite{farmer1983}, meaning that it has non-trivial geometry on all scales. Fractals can be categorized by their dimension $D$, which describes how a set's geometrical features change, depending on the scale at which we observe it. The original definition of $D$ goes back to Felix Hausdorff and can conveniently be approximated as the limit $\epsilon\to 0$  of $-\ln N(\epsilon)/\ln(\epsilon)$, where $N(\epsilon)$ is the minimum number of $M$-dimensional hypercubes of side-length $\epsilon$ needed to cover a set $S\in\mathbb{R}^{M}$ (cf. Ref. \onlinecite{mandelbrot1983}).\par 
In the context of chaotic dynamical systems, we can define the attractor dimension, henceforth denoted by $D_1(A)$, as an indicator of randomness and predictability: Consider the random process of observing the system's position in phase space up to an accuracy of $\epsilon$. $D_1$ is defined such that, for sufficiently high accuracies, the entropy $H_\epsilon$ of this process (i.e. the expected information from a single observation, as defined in Ref.\,\onlinecite{shannon2001})
obeys the scaling law $H_\epsilon\propto -D_1\ln(\epsilon)$. The dimension $D_1$ is therefore sometimes referred to as \textit{information dimension}. It can be shown \cite{ott2002} that this is an appropriate generalization to the purely geometric notion of scaling features. 
Furthermore, let $\lambda_1>\hdots>\lambda_N$ be the Lyapunov numbers describing our system's error evolution. According to the Kaplan-Yorke-conjecture \cite{kaplan1979}, a small volume in phase space extending in the first $j$ directions (ordered by the $\lambda_i$) will expand if $j<D_1$ and otherwise contract, $D_1$ being the exact interpolated value in between \cite{karimi2010}. 
The attractor dimension is thus closely related to the error growth and thereby predictability of the system -- intuitively, trajectories on high dimensional sets may evolve in many directions, leading to challenging forecast problems.\par
The non-trivial task of estimating $D_1$ has been significantly facilitated by the recent realization \cite{collet2001, freitas2010,faranda2011,lucarini2012a,lucarini2012b} that $D_1$'s local counterpart $\locdim$, defined such that its expectation value $E[\locdim]$ equals $D_1$, governs the probabilities of extremely close recurrences in phase space.  For a complete account of the theoretical background, the reader is referred to Ref.\,\onlinecite{lucarini2016}. Using these results, the so-called \textit{local dimension} $\locdim$ (defined by equation \ref{eq:Dp}) can be estimated in three simple steps: 1.) Observe a long orbit $x_1,\hdots , x_n$ of the system. 2.) For some chosen point along the orbit, calculate the distance to all other $x$. 3.) Select the extremely small distances. The probability distribution of these extremes depends only on $\locdim$, which can subsequently be inferred via  standard techniques of parameter estimation.\par
Several recent studies have begun to exploit this new possibility, working under the assumption that $\locdim$ plays a similar role for the predictability of individual states as $D_1$ for the average properties of the system. Ref.\,\onlinecite{faranda2017} estimated $\locdim$ from recurrences in NCEP-reanalysis sea level pressure data over the North Atlantic region. These authors report on a robust connection between a state's local dimension and the current phase of the North Atlantic Oscillation (NAO), noting that exceptionally small values of $\locdim$ usually correspond to strong NAO$^+$ events. Analyzing the uncertainty of ensemble predictions from initial conditions of different dimensionality, they were furthermore able to empirically support the hypothesis that $\locdim$ is a useful indicator of predictability. In a similar study, Ref.\,\onlinecite{messori2017} highlighted the possibility of employing local dynamical indicators like $\locdim$ (estimated there from upper-level geopotential) as predictors of European temperature extremes. Further studies suggest applications for model evaluation and detection of climate change signals \cite{rodrigues2017}, attempt to estimate $\locdim$ from various atmospheric fields and investigate the influence of seasonality \cite{farandadynamical}\par
In this study, we aim to reinforce and expand on these results in two main ways. Firstly, the connection between local dimensions and dominant modes of variability has so far only been observed in reanalyses. While such data-sets do represent our best estimate of the true atmospheric trajectory, they differ from the ideal case for which our methodology is designed: The dynamics are not stationary due to the presence of seasonal cycles, data assimilation effectively introduces a stochastic component and the length of available trajectories is inherently limited. It is thus important to ensure that previous observations were not the product of violated assumptions or random chance. To that end, we perform experiments with the PUMA model \cite{fraedrich1998}. In this simplified representation of the atmosphere, random external forcing is absent, stationarity can be achieved and simulations over multiple centuries are readily available.\par
Secondly, we propose an alternative estimation procedure based on an analytic maximum likelihood estimator of $\locdim$, which is expected to reduce biases and enables a straightforward test of agreement between theoretically expected recurrence behavior and observations. Using this result, we are furthermore able to identify a potential source of error in estimates of $\locdim(x_i)$, related to the direct predecessors and successors of $x_i$. It is argued, that the newly proposed methodology is almost always preferable to the available numerical alternatives.\par
In a potentially interesting aside (section \ref{sub:IDx}), we point out that $\locdim$ is closely related to the concept of intrinsic dimensionality, which has recently been introduced in the field of similarity search \cite{houle2015}. In particular, Ref.\,\onlinecite{levina2005} and Ref.\,\onlinecite{amsaleg2014} have independently arrived at the exact same analytical estimator we propose here, albeit with completely different uses and interpretations in mind. We will show the connection between the two dimensions and hint at possible applications.\par
The paper is organized as follows: Relevant assumptions and definitions are introduced in the beginning of section \ref{sec:meth}. We describe the procedures employed by previous authors and develop our altered approach. The connection to the works of Ref.\,\onlinecite{houle2015} and Ref.\,\onlinecite{amsaleg2014} is outlined as well. 
Section \ref{sec:data} contains a description of the used PUMA configuration, all experimental results are collected in section \ref{sec:res1}. We conclude with a discussion in section \ref{sec:dis}.

\section{Methods}\label{sec:meth}
For our purposes, a measure preserving dynamical system is given by its phase-space $\Omega$ and a map $T:\Omega\to\Omega$ determining the time evolution from  $x_n\in\Omega$ to $x_{n+1}=T(x_n)$. We consider the points $x_i$ as realizations of a random variable $X$, assuming that inexact knowledge of initial conditions prevents us from knowing the system's position at time $n$ exactly. The chance of observing $X$ within some subset of $\Omega$ is then given by the natural probability measure $\mu:\mathcal{B}(\Omega)\to[0,1]$, where $\mathcal{B}(\Omega)$ is the Borel algebra on $\Omega$.
The requirement that $\mu$ be preserved under the time evolution, i.e. $\mu(T^{-1}S)=\mu(S)$ for all $S\in\mathcal{B}(\Omega)$, implies that every set $S$ with non-zero measure must be revisited arbitrarily closely by $\mu$-almost all trajectories originating within $S$. This statement, known as the Poincar\'{e} recurrence theorem \cite{poincare1890}, forms the basis of our methodology: For some chosen point $x_i$ along the orbit, the random variable $R_i:=dist(x_i,X)$ takes on values arbitrarily close to zero. We study these small distances by transforming them into positive extremes of some strictly monotonously decaying function $g(R_i)$, as is a standard technique in extreme value (EV-) statistics \cite{coles2001}. The function $g$ is sometimes referred to as a \textit{distance observable}. To define the extremes, we choose a peak over threshold (POT) approach, thus considering a recurrence extreme if $g$ exceeds some high threshold $u$. If we dispose of an i.i.d. sample of $g(R_i)$, classical results of EV-statistics \cite{pickands1975,balkema1974} guarantee that, in the limit of large $u$, the conditional tail distribution 
\begin{align}
	H_i(z):=Prob\left[g(R_i)-u>z\,|\,g(R_i)>u\right]
\end{align}
must be either degenerate or a member of the Generalized Pareto (GP) family, i.e.
\begin{align}
	\exists \,\sigma,\xi:\hspace{5pt} H(z)&=GP(z,\sigma,\xi)\nonumber\\
	&=1-\left[1+\xi(z-u)/\sigma\right]_{+}^{-1/\xi}\comgood
\end{align}
where we have used the notation $\psi_+:=max(\psi,0)$.
Furthermore, conditions that guarantee the existence of this limit distribution are available.
While it has long since been realized that the independence-requirement can be substantially relaxed \cite{coles2001}, the development of an EV-theory for perfectly deterministic dynamical systems is a relatively recent endeavor, the history of which is outlined in Ref.\,\onlinecite{lucarini2016}. In particular, Ref.\,\onlinecite{lucarini2012b} considered dynamical systems for which the attractor has a well-defined local dimension $\locdim(x_i):=\locdim_i$, i.e.
\begin{align}
	\mu(B_\epsilon(x_i))\propto \epsilon^{\locdim_i}\comgood\label{eq:Dp}
\end{align}
where $B_\epsilon(x_i)$ is the open $\epsilon$-ball around $x_i$. It is straightforward to show that this implies the entropy scaling $H_\epsilon=E[\mu(B_\epsilon)]\sim-E[d]\ln(\epsilon)$ mentioned in the introduction. Furthermore, the authors cited above found that $g$ then obeys the following extremal law: 
\begin{align}
	H_i(z)=\left[g^{-1}(z+u)/g^{-1}(u)\right]^{\locdim_i}\dotgood\label{eq:cdf}
\end{align}
For some simple choices of $g$, it can be shown \cite{lucarini2012b} that $H$ is a member of the GP-family, despite dependence in the series (i.e. despite the fact that the classical theorem may not apply). Furthermore, the two GP-parameters $\sigma$ and $\xi$ depend only on $\locdim$ and our choice of $g$. In the convenient special case $g(R_i)=-\log(R_i)$, which was employed in Refs. \onlinecite{faranda2017, messori2017, rodrigues2017}, the parameters become $\xi=0$ and $\sigma=1/\locdim$. Using that result, those authors have estimated $\locdim$ as follows: 1.) Calculate series of $g(R_i)$. 2.) Choose the threshold $u$ as some high quantile of $g$'s empirical distribution. 3.) Fit a GP-distribution to the threshold exceedances and estimate $\locdim$ as the inverse of $\sigma$.\par

\subsection{Analytical estimator}\label{sub:news}
While previous authors have thus exploited the relationship between $\locdim$ and the GP scale-parameter $\sigma$, we will now demonstrate how the second estimated parameter $\xi$ can be used to test the agreement between theory and experimental results, even if the true value of $\locdim$ is unknown: In addition to the fit with two free parameters described above, we perform a second optimization where the value of $\xi$ is fixed at zero. Let $\ell$ be the optimized log-likelihood of the former, and $\ell_0$ be the corresponding value for the latter fit. If $\xi\equiv 0$ is indeed optimal, as it must be if $\locdim$ is well defined, Wilks' theorem \cite{wilks1938} states that the quantity $L:=2(\ell-\ell_0)$ follows a $\chi^2$-distribution with one degree of freedom. We can thus perform a likelihood ratio test \cite{coles2001} of the hypothesis that the statistics of extreme recurrences are entirely determined by $\locdim$ in the sense detailed above.\par
Besides its role in this goodness-of-fit test, the constrained maximum likelihood estimator with fixed $\xi=0$ has further welcome properties: We can derive analytic expressions for its value as well as its uncertainty and they are independent of our choice of $g$. To see this, we simply note that using the constraints a priori is equivalent to the direct optimization of the likelihood resulting from (\ref{eq:cdf}). The relevant probability density is
\begin{align}
	-\frac{\text{d}}{\text{d}z}H_i(z)=-H_i(z)\,\locdim_i\, \frac{\text{d}}{\text{d}z} \ln\left(g^{-1}(z+u)\right)\comgood
\end{align}
where the last factor on the right hand side is independent of $\locdim_i$. The threshold $u$ is set to some high quantile $\tau$ of $g(R_i)$'s empirical distribution, meaning that the positions of threshold exceedances are independent of $g$. After a few elementary steps, we find that the likelihood resulting from this density is optimized at 
\begin{align}
	\hat{\locdim}_i=\left(\overline{\log u_R-\log R_i}\right)^{-1}\comgood\label{eq:dhat}
\end{align}
where we have set $u_R:=g^{-1}(u)$, and the horizontal bar denotes the sample mean over all $r_i<u_R$. The estimator is asymptotically normally distributed, its standard deviation  satisfying
\begin{align}
	sd(\hat{\locdim_i})=\hat{\locdim_i}/\sqrt{m(1-\tau)}\comgood\label{eq:unc}
\end{align}
where $m$ is the number of points along the orbit. This result can serve as a rough estimate of the accuracy with which $\locdim_i$ is estimated, noting that (\ref{eq:unc}) only exactly holds in the limit of infinitely close recurrences.\par
Equation \ref{eq:dhat} arises very naturally in the special case $g(R_i)=-\log(R_i)$, where our estimate equals the inverse mean threshold exceedance of that observable (cf. Ref.\,\onlinecite{pons2018} who exploit this fact in a similar context). The derivation above however demonstrates that this estimator does not depend on our choice of $g$ at all, instead applying to all valid distance observables.


\subsection{Effects of direct temporal neighbours}\label{sub:neigh}
Before we can put the suggested methodology to the test, we must consider an important technical detail. While chaotic dynamics justify the description of returns to $B_\epsilon(x_i)$ as a stochastic process, this treatment appears unsuitable for the direct predecessors and successors of $x_i$ (i.e. $\hdots,x_{i-2},x_{i-1},x_{i+1},x_{i+2},\hdots$). It is easy to see how the effect of these direct neighbors in time may dominate our results in the case of finite data sets. Suppose we estimate $\locdim_i$ from a collection of points $x_{i+j},x_{i+j+1},\hdots$ such that direct neighbors are excluded. Note that the derivation of our estimates relies on no assumptions about the time-difference $j$.
We should therefore expect that $\hat{\locdim_i}\approx\locdim_i$, provided that $m$ is large. This should still hold if we set $j=1$, but since the distances $r_{i+1},\hdots,r_{i+j-1}$ will be unusually small (resulting from only a few iterations of $T$), the term $-\overline{\log R_i}$ in (\ref{eq:dhat}) is increased and $\hat{\locdim}_i$ is lowered, thus introducing a negative bias. Note that this argument will only hold, if recurrences are in fact rare and the immediate predecessors and successors are the points closest to $x_i$. For low-dimensional systems (such as those studied in Ref.\,\onlinecite{lucarini2012b}) and long trajectories, we should expect no relevant impact since numerous true recurrences will easily outweigh the temporal neighbors (see black dots in figure \ref{fig:Dpoftau_a}). It is furthermore not clear, whether our heuristic argument should apply to the unconstrained estimators of $d$, where the second degree of freedom in the fitting procedure could allow for unexpected results.

\subsection{Further dimensions and their estimators}\label{sub:IDx}
The question of determining a set's unknown dimension has numerous applications outside of geophysics. One interesting example is the field of similarity search, which is concerned with the evaluation of distances between points in a database that possesses no natural ordering \cite{chavez2001}. Dimension-estimates of such point-clouds, both globally and locally, characterize the discriminability of points and are needed to assess the possibility of dimension reduction. In a recent publication, Ref.\,\onlinecite{houle2015} have introduced the so-called \textit{continuous intrinsic dimension} $\intdim_i(r)$, which measures how the probability of finding points near $x_i$ scales with the search radius $r$. In appendix \ref{app:intdim}, we give the explicit definition of $\intdim$ and prove that $\locdim=\lim\limits_{r\to 0}\intdim(r)$. A very similar result has already been proven by Ref.\,\onlinecite{romano2016}, who recognized that $E[\intdim(0)]=D_1$. More generally, these authors show that all R\'{e}nyi-Dimensions $D_q$ (i.e. scaling coefficients of generalized entropy, see Ref.\,\onlinecite{renyi1961}) can be expressed in terms of $\intdim(0)$ and thus in terms of $\locdim$. Beyond the inherent noteworthiness of these observations, we mention them in the present context for three main reasons: 
Firstly, Ref.\,\onlinecite{amsaleg2014} have used extreme value theory to estimate $\intdim(0)$. Assuming the validity of the classical result (i.e. convergence to a GP-distribution), they arrived exactly at the estimator given in (\ref{eq:dhat}). These authors furthermore performed experiments with several non-ML estimators (including probability weighted moments and regularly varying functions). Their results suggest that the ML-estimator performs no worse than its competitors -- an observation which may well be transferable to our application.\par
Secondly, Ref.\,\onlinecite{faranda2018correlation} have recently proven that the probabilities of extremely close encounters between two independent trajectories of a dynamical system are governed by the correlation dimension $D_{q=2}$, which can thus be inferred from observations. It is clear that their result should be closely related to that of Ref.\,\onlinecite{romano2016}. While further work beyond the scope of our present investigation is needed to clarify that connection, it could eventually lead to additional ways of independently confirming the validity of our experimental results.\par 
Thirdly, it turns out that (\ref{eq:dhat}) has been used to estimate local dimensionality even before the introduction of $\intdim$: Ref.\,\onlinecite{levina2005} start with the assumption that the density of points close to $x_i$ is nearly uniform and the number of points hitting $B_r(x_i)$ corresponds to a Poisson point process. Omitting an explicit definition entirely, they arrive at (\ref{eq:dhat}) as an estimate of the dimension near $x_i$. It is not obvious that their approximation of the density within $B_r(x_i)$ is equivalent to the definition of $\intdim$, but the existence of a common estimator suggests that they should be closely related. These authors also strongly advocate the use of the ML-estimator over other popular methods in their field.\par 
The results of Ref.\,\onlinecite{levina2005} are furthermore relevant to us, because their approximation to the probability density near $x_i$ can be used to derive a second, non-ML estimator for the global attractor dimension $\overline{D}$: Making the same Poisson assumption, Ref.\,\onlinecite{pettis1979} have developed an iterative procedure which determines $\overline{D}$ via linear regression of the number of near neighbours $\log(k_r)$ against their mean distance $\log(\overline{r})$. For details on the exact procedure, we refer to the original publication. Noting that this ``neighbourhood''-estimator yields correct results in simple systems of known dimension (preliminary experiments, not shown) and requires the exact same distance data as (\ref{eq:dhat}), we employ it for comparison in section \ref{sub:locdim}.

\section{Data}\label{sec:data}
If preservation of $\mu$ and the existence of $\locdim$ are assumed, the framework presented above can be applied to a wide variety of systems, from low dimensional ``toy''-models such as those tested by Ref.\,\onlinecite{lucarini2012b} to realistic atmospheric trajectories --  with an important restriction: We implicitly rely on the idea that time evolution can in fact be represented as $x_{n+1}=T(x_n)$, meaning that one state is always mapped to the same successor. Atmospheric reanalysis data sets, which incorporate observations, appear to violate this assumption in at least two ways: Seasonal cycles, driven by external forcing introduce an explicit time-dependence to $T$ and the process of data assimilation constitutes a stochastic component, since the value of the observations is not determined by the dynamics themselves. While some progress towards extending the theory presented in section \ref{sec:meth} to the stochastically perturbed case has already been made\cite{faranda2013extreme,faranda2014extreme}, we consider the purely deterministic case here. To that end, we employ the PUMA model \cite{fraedrich1998}, which combines the dynamical core of the ECHAM climate models with rudimentary parameterizations of thermodynamics and friction similar to those proposed by Ref.\,\onlinecite{held1994} and contains no stochastic terms. Orography was activated to allow for some degree of comparability with the reanalysis data considered by previous authors. We have adapted the Held and Suarez \cite{held1994} relaxation temperature profile (natively implemented in PUMA only for aqua-planets) for simulations with realistic orography by replacing the constant reference pressure $p_0$ by the true average surface pressure. 
The only randomness in our experiment stems from the initial conditions. We enforce stationarity by deactivating the seasonal cycle. The system is left in eternal winter (specifically with the thermal forcing for the first of January), which was selected because Ref.\,\onlinecite{faranda2017} report that the minima of $\locdim$, supposedly related to the NAO, occur predominantly in winter. Our analysis is based on the daily instantaneous output over a period of 1000 years with spectral resolution T42L10 (triangular truncation at wavenumber 42, corresponding to a Gaussian grid with $128\times 64$ grid points, ten vertical levels).\par
In this setup, there appears to be no relevant fundamental difference between the PUMA model and simple low dimensional attractors like the original system considered by Ref.\,\onlinecite{Lorenz1963}, even though in the latter case, the map $T$ is somewhat less complex. Nevertheless, we cannot treat the two systems equally: In order to calculate the distance between two complete atmospheric state vectors, we must decide how to weigh the different physical quantities. 
While some possible solutions are available in the literature (for example energy norms, as used by Ref.\,\onlinecite{hense1995}), the question largely remains open at this point. Even if this complication was resolved, we would be left with the problem of excessive computational costs. In this study, we therefore follow the approach of previous authors and truncate the state vectors to a single field in a limited geographical region. Specifically, we choose the same domain as Ref.\,\onlinecite{faranda2017}, i.e. sea level pressure in the North Atlantic region (22.5$^\circ$N - 70$^\circ$N, 80$^\circ$W - 50$^\circ$E), thus enabling a direct comparison.\par 
It is important to note that the truncation constitutes a substantial departure from the theoretical assumptions, which almost surely implies that we cannot obtain an accurate estimate of $D_1$ for the complete system: We could either treat the effect as a potentially huge error in the calculation of distances or consider the truncated state as a stochastic system (driven by the unobserved variables). A rigorous treatment of this problem is beyond the scope of our investigation. Results of previous authors suggest that $\locdim$ may in either case provide useful information about the considered subsystem.

\section{Results}\label{sec:res1}

As mentioned in the introduction, it is our first objective to reinforce and confirm results that were previously obtained with comparably short trajectories and in the less idealized setting of reanalysis data sets. In a first step, we check whether the recurrence behavior of the PUMA simulation appears to be governed by a well defined local dimension. 
The performance of our analytical estimator has to be tested as well.
Following this, we investigate the connection between exceptional values of $\locdim$ and dominant patterns of circulation. 

\subsection{Estimates of the local dimension}\label{sub:locdim}
We calculate the local dimension $\locdim$ for 10000 randomly selected points along the trajectory of our PUMA run using the two maximum likelihood estimators and the ``neighborhood'' approach of Ref.\,\onlinecite{pettis1979}. To assess the possible convergence to some limit value, we vary the threshold quantile $\tau$ between 99\,\% and 99.9\,\%. In order to keep the results of the likelihood ratio tests comparable between thresholds, each estimate is based on 360 randomly sampled threshold exceedances, corresponding via (\ref{eq:unc}) to roughly 5\,\% relative error in the asymptotic limit.
\begin{figure}[hb]
		\includegraphics[width=.49\textwidth]{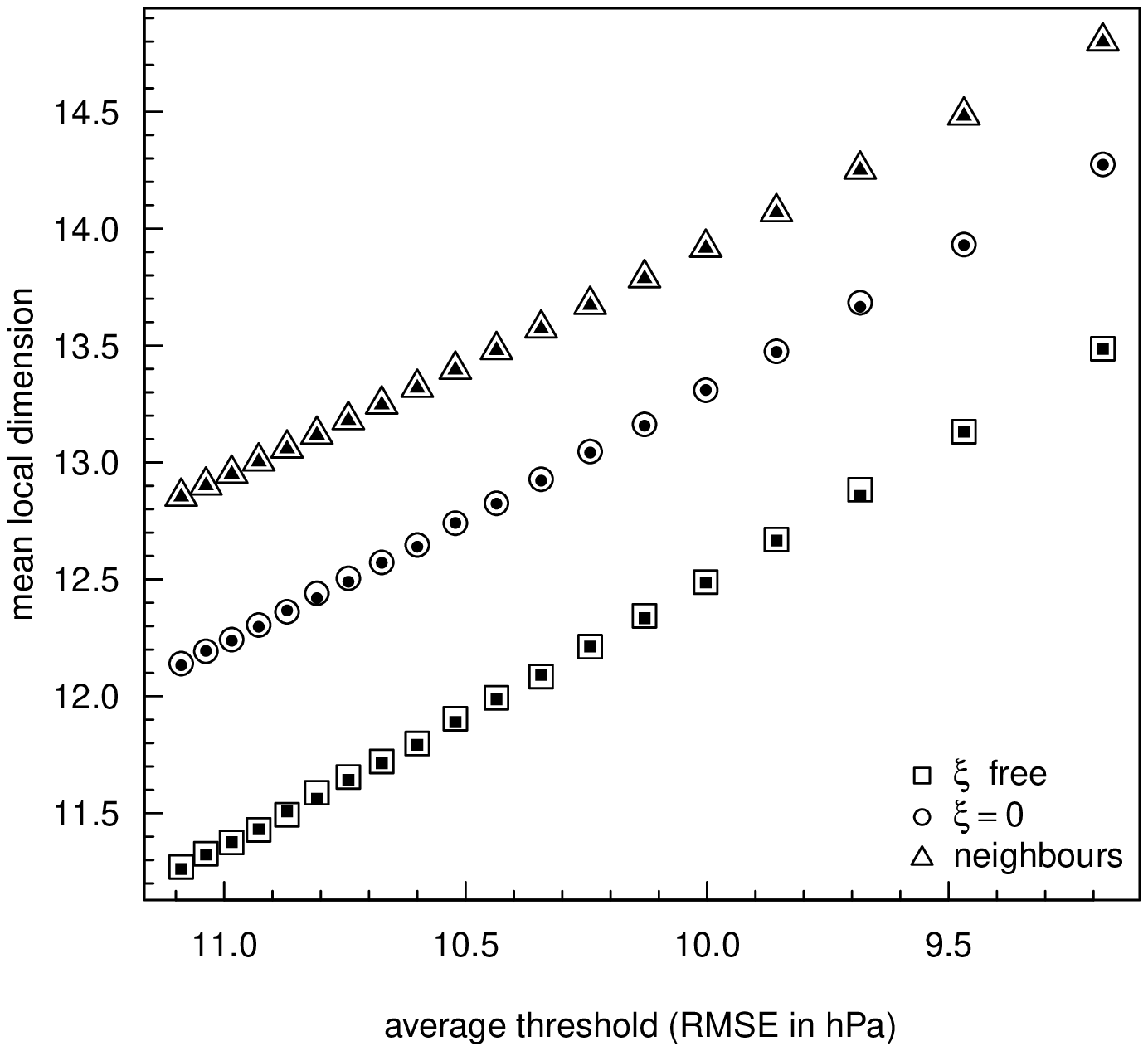}
	\caption{Estimates of mean local dimension as a function of the mean threshold, written here as root mean square difference, i.e. $\overline{u_R}/\sqrt{\#grid\,\,points}$. Threshold quantiles range from $99\,\%$ to $99.9\,\%$. Small black symbols indicate results obtained without eliminating direct temporal neighbours.}\label{fig:Dpoftau_a}
		\includegraphics[width=.49\textwidth]{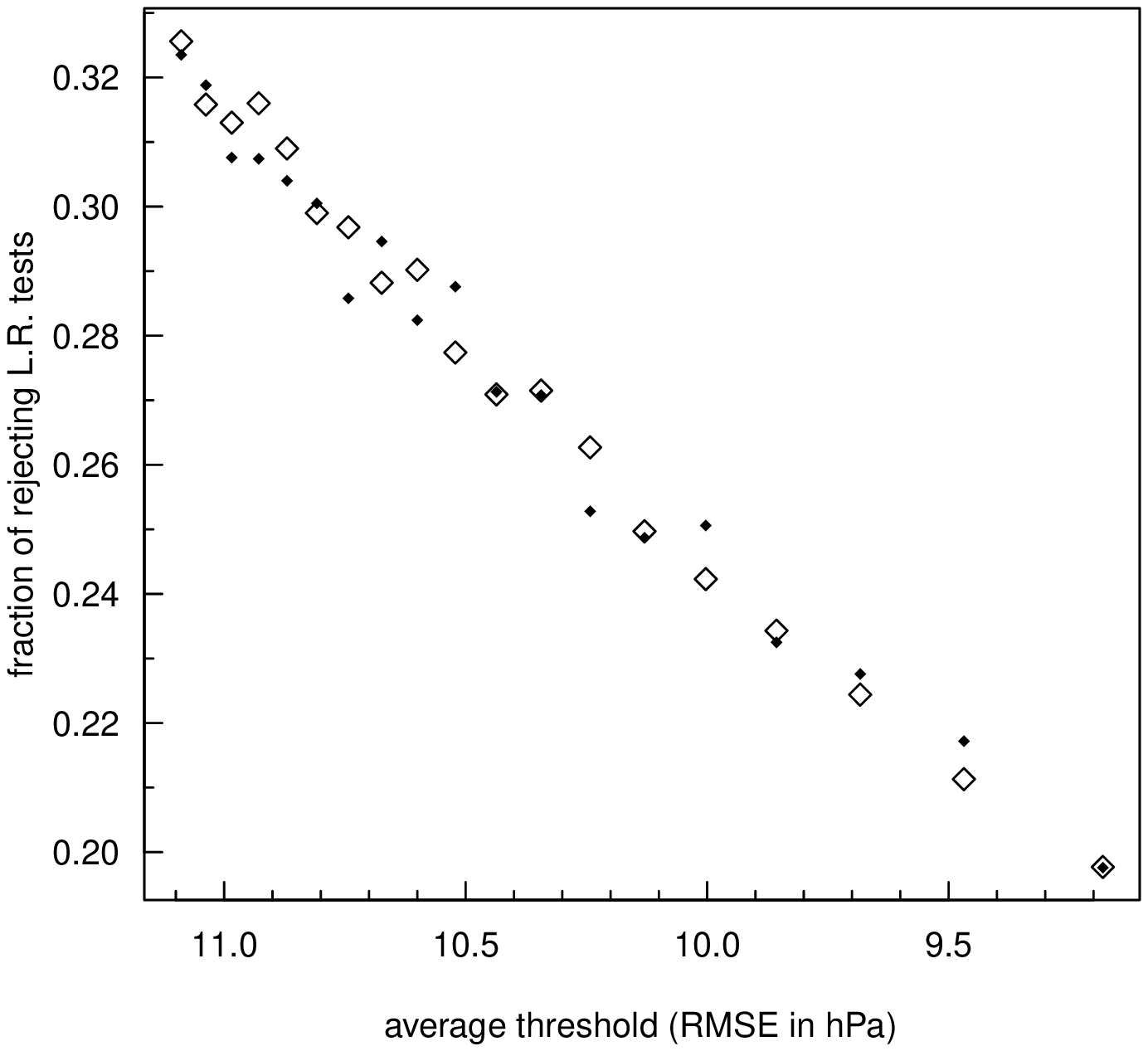}
	\caption{Fraction of cases where the hypothesis $\xi\equiv 0$ is rejected at $p=5\,\%$ as a function of the mean threshold.}
	\label{fig:Dpoftau_b}
\end{figure}
For visualization in figures \ref{fig:Dpoftau_a} and \ref{fig:Dpoftau_b}, we divide the mean threshold $\overline{u_R}$ by the square root of the number of grid points (in this case $\sqrt{782}$), thus translating it into a more interpretable quantity, namely the root mean square difference of two pressure fields. We have checked that all of the GPD-fits are reasonably good (fewer than 5\,\% of Anderson-Darling tests reject the hypothesis that the observed sample is drawn from the fitted distribution at $p=5\,\%$).
All three estimates of the mean attractor dimension (Figure \ref{fig:Dpoftau_a}) rise monotonically as the threshold is increased, each remaining far below the dimension of the ambient space (i.e. 782). Their relative order (neighborhood estimate $>$ constrained ML-estimate $>$ free ML-estimate) is independent of the chosen threshold, none of the three curves exhibit a slowing growth rate, thus giving no immediate indication of convergence. 
If we assume for a moment that the local dimensions do give us an estimate of the complete system's global dimension (with errors introduced by the truncation of the state vector), we can compare the numbers in figure \ref{fig:Dpoftau_a} to the results of Ref.\,\onlinecite{cruz2018}. These authors used covariant Lyapunov vectors to estimate the information dimension of PUMA (T42L10) via the Kaplan-Yorke formula, finding values on the order of 100. While their model-configuration is somewhat different from ours (no orography, different background temperature field), this nonetheless confirms that no reliable estimate of the global attractor dimension can be obtained via naive estimation of $\locdim$ from a trajectory of 1000 years. There is thus little hope of achieving such an estimate in the context of reanalysis data, as was already recognized by Ref.\,\onlinecite{faranda2017}.\par

The fraction of significantly non-zero shape parameters (figure \ref{fig:Dpoftau_b})  shows that the behavior is nonetheless slowly approaching the expected limit: While the absolute number of L.R.-tests rejecting the hypothesis $\xi\equiv 0$ is of course determined by the sample size, the constant negative trend clearly indicates increasing agreement with theory. Simultaneously, the average estimated shape parameters $\xi$ approach zero from below, changing from approximately $-0.08$ to $-0.06$ across the tested thresholds (not shown). The differences between the three estimates of $D_1$ are very slowly (but monotonously) decreasing as well.
We have furthermore checked that the correlations between $\locdim(\xi=0)$ and $\locdim(\xi\neq 0)$ are very high ($>0.9$) for all values of $\tau$. This implies that the choice of estimation procedure makes little difference to the kind of analysis carried out in section \ref{sub:circ} below. In the context of our present investigation, the same can be said about the effect of direct temporal neighbors: Neither the estimated dimensions nor the rate of rejecting L.R.-tests change substantially if we include $x_i$'s immediate predecessors and successors (see small black symbols in figure \ref{fig:Dpoftau_a} and \ref{fig:Dpoftau_b}).\par

\begin{figure}[!t]
	\centering
	\includegraphics[width=0.49\textwidth]{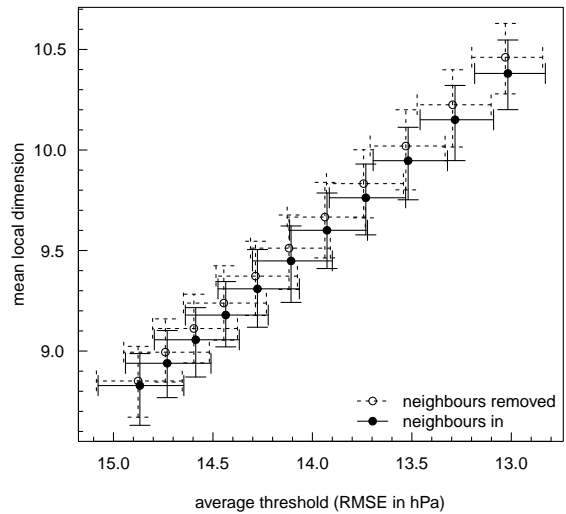}
	\caption{Constrained estimators of $D_1$, calculated from 5\,year-sections of the original orbit. Threshold quantiles were varied between 85\,\% and 95\,\% all estimates are based on 90 randomly sampled recurrences. Points indicate mean over all sections, error bars denote 95\,\%-ranges.}
	\label{fig:Dpoftau_30d}
\end{figure}

Following the argument presented in section \ref{sub:neigh}, we expect neighbors in time become relevant for shorter trajectories, as the number of actual recurrences is reduced. By removing the seasonal cycle, we have effectively increased the number of possible recurrences by approximately one order of magnitude: Visual inspection of typical time series of distances in phase space for an orbit with seasonality (namely the NCEP reanalysis data-set used by Ref.\,\onlinecite{faranda2017}, not shown) confirms that recurrences are mostly constrained to dates within the same phase of the cycle. We can roughly simulate this by cutting the original long trajectory into sections of five years. This corresponds to one month from each of sixty years, i.e. the approximate number of ``recurrence-candidates'' in the NCEP data set. In this set-up, the neighbors introduce a negative bias, which grows with increasing threshold (see figure \ref{fig:Dpoftau_30d}) and is eventually no longer negligible compared to the variability of the estimated means. The correlation between estimates of $\locdim$ with and without temporal neighbors lies around 0.81  (independent of threshold) meaning that the added error is mostly, but not entirely systematic.

\subsection{Connection to circulation patterns}\label{sub:circ}
A second major finding of Refs. \onlinecite{faranda2017} and \onlinecite{messori2017} is the connection between extremes of the local dimension and certain well known patterns of circulation. In particular, these authors found that the positive phase of the North Atlantic Oscillation (NAO) correlates with exceptionally low values of $\locdim$. \par 
To identify the major patterns of variability in our simulation, we apply a principal component analysis. 
\begin{figure*}
	\includegraphics[width=\textwidth]{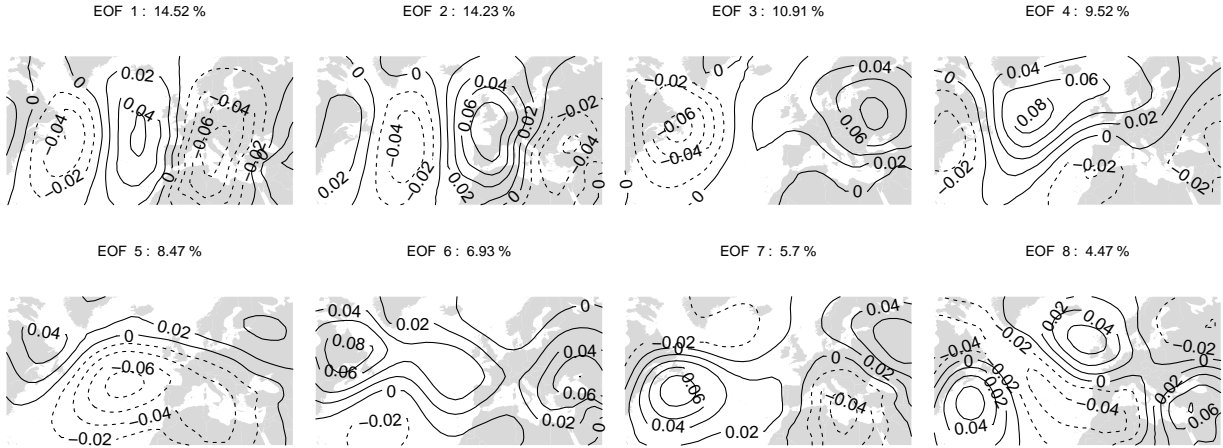}
	\caption{Leading EOFs of daily instantaneous sea level pressure in PUMA. Percentages indicate explained variance.}
	\label{fig:EOFs}
\end{figure*}
Figure \ref{fig:EOFs} shows the first eight empirical orthogonal functions (EOFs), i.e. the estimated eigenvectors of the spatial covariance matrix. The two dominant vectors correspond to a Rossby wave propagating eastwards through the domain. They are followed by various wavenumber one and two patterns, none of which look particularly reminiscent of the real-world North Atlantic dipole. These higher-order vectors are notoriously hard to interpret since their form is increasingly determined by the forced orthogonality to all lower-order EOFs. Without additional information, it appears impossible to decide which, if any, of these patterns might correspond a real dynamical phenomenon. We will now demonstrate how an analysis of recurrence behavior, using the local dimension $\locdim$, may in fact provide answers to such questions. The results presented hereafter are based on the constrained estimates, using the highest tested threshold-quantile ($\tau=99.9\,\%$). We have checked that neither lifting the constraint nor slightly lowering the thresholds leads to any substantial qualitative changes in this context.\par  
\begin{figure*}[!t]
	\centering
	\includegraphics[width=\textwidth]{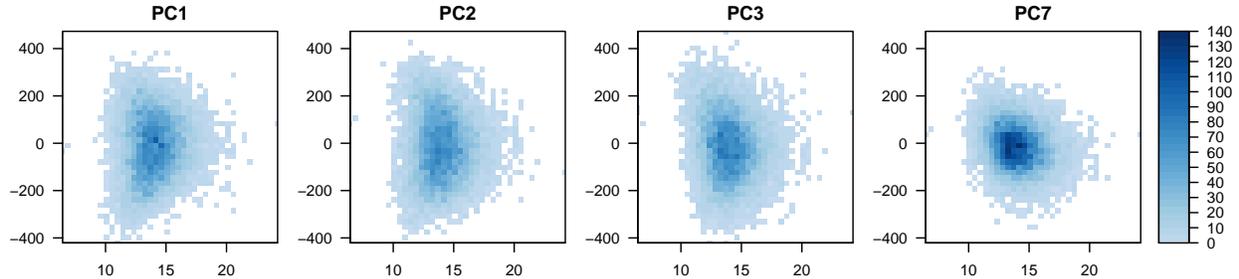}
	\caption{Two dimensional histograms of $\locdim$ (x-axis) and some of the leading principal components (y-axis). Color indicates the number of days within the respective bins.}
	\label{fig:EVs_and_cor}
\end{figure*}
\begin{figure}[!b]
	\centering
	\includegraphics[width=0.49\textwidth]{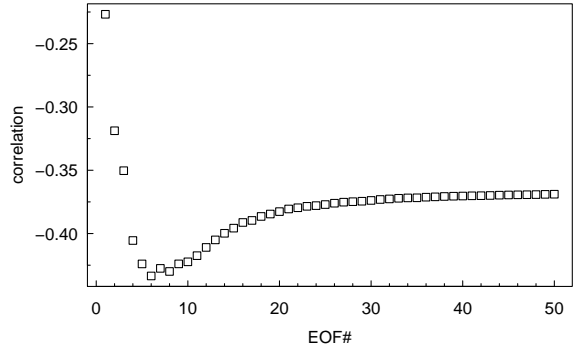}
	\caption{Correlation between $\locdim$ and the sum over the first $m$ principal components squared ($m$ on the $x$-axis).}
	\label{fig:PCcors}
\end{figure}

Computing the correlations between $\locdim$ and the principal components (henceforth $PC$s), we find that several of the seemingly relevant patterns, particularly EOFs three and seven, have weak but non-negligible connections to $\locdim$ (correlations close to $0.15$ and $0.2$, respectively, significant at $p=1\,\%$ according to a Pearson-test). The corresponding bivariate histograms of $\locdim$ and the principal components (figure \ref{fig:EVs_and_cor}) confirm that particularly small local dimensions tend to coincide with the positive phases of EOF3 and EOF7. While the specific phase of the Rossby-wave (represented by the two leading vectors) is uncorrelated with $\locdim$, we observe that strong amplitudes of this pattern only occur when the local dimension is relatively low. In fact, we find $cor(PC1^2+PC2^2,\locdim)\approx -0.32$. Figure \ref{fig:PCcors} reveals that this negative correlation continues to decrease as we add the squares of further components, reaching the minimum at $PC6$ before approaching a limit of approximately $-0.37$.
To analyse the nature of high- and low-dimensional configurations in detail, we begin by considering the average pressure fields on days with extreme values of $\locdim$.

\begin{figure}[!b]
		\includegraphics[width=.49\textwidth]{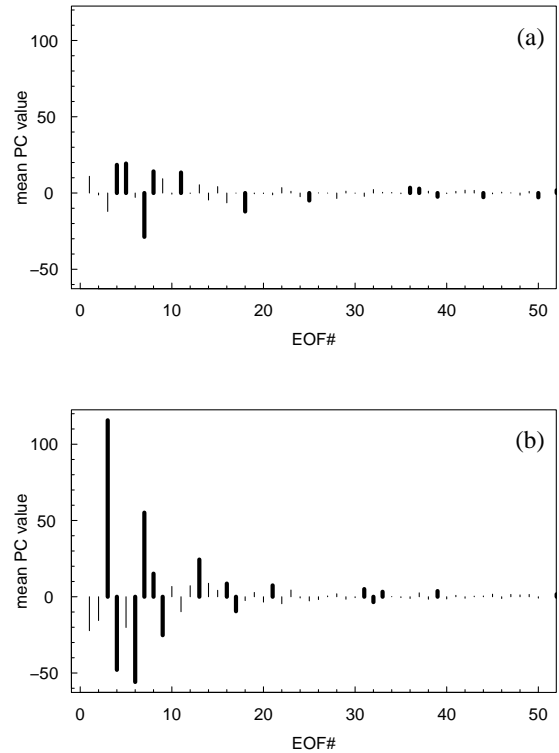}
	\caption{First 50 principal components of pressure fields, averaged over days with local dimensions above the $98\,\%$-quantile (a) and below the $2\,\%$-quantile (b). Thick lines indicate significantly non-zero mean according to $t$-tests at $p=1\,\%$.}
	\label{fig:PC_extreme}
\end{figure}

When $\locdim$ falls into the lower 2\,\% of its distribution across the attractor, the mean state is indeed dominated by the positive phase of EOF three, with some smaller significant contributions from higher order vectors (figure \ref{fig:PC_extreme}\,b). We have checked that this result is insensitive to the arbitrarily chosen 2\,\%-threshold. As expected, the moving Rossby-wave averages out since neither individual phase has preferred values of $\locdim$. 
\begin{figure*}[!t]
	\centering
	\includegraphics[width=\textwidth]{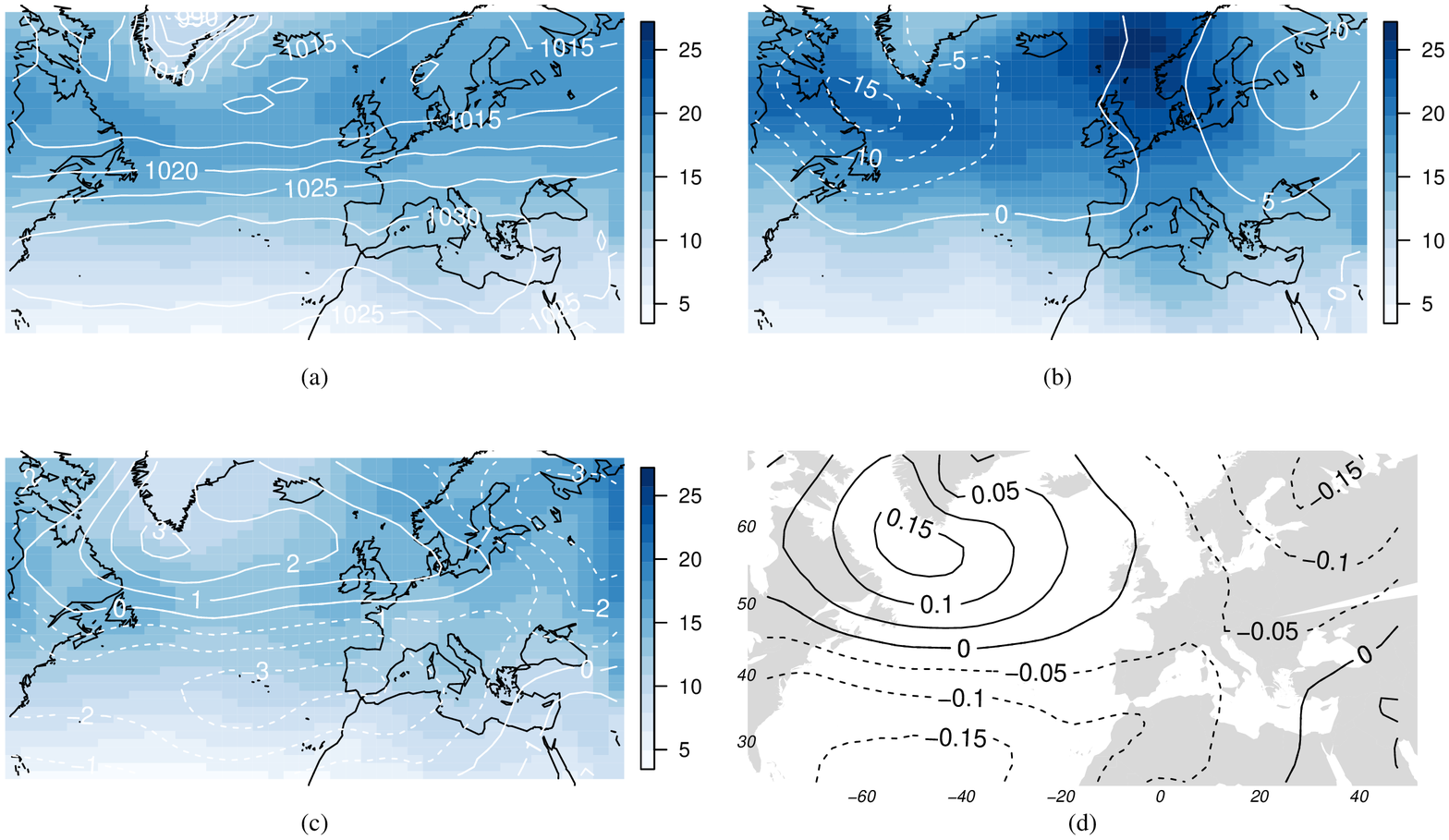}
	\caption{Climatological mean (contours) and standard deviation (shading) of sea level pressure, calculated over all 10000 selected times (a), as well as anomalies over times with $\locdim$ below the 2\,\%-quantile (b) and times with $\locdim$ above the 98\,\%-quantile (c). Panel (d) shows the gridpoint-wise correlations between pressure and $\locdim$.}
	\label{fig:PC_extreme_geo}
\end{figure*}

The nature of these low-dimensional states becomes more obvious when we transform the composite back into geographical coordinates. The climatological mean field (figure \ref{fig:PC_extreme_geo}\,a) is nearly zonally symmetrical  with high pressure in the South and lower pressure in the North of the domain. Local extremes near Iceland and the Azores, as typically seen in real world climatologies for the winter season, are notably absent. This dipole structure does however appear in the mean anomaly over days with low values of $\locdim$ (figure \ref{fig:PC_extreme_geo}\,b): Meridional pressure gradients are enhanced over the Atlantic and greatly reduced over continental Europe. The gridpoint-wise correlation between $\locdim$ and sea level pressure (figure \ref{fig:PC_extreme_geo}\,d) confirms that this pattern is not only present in the extreme tail of $\locdim$'s distribution, but observable in the complete data set. While very little of the pressure variance is explicable by $\locdim$ itself, a decrease in dimension does typically coincide with low pressure south of Greenland and high pressure near the Azores. We furthermore note that the temporal standard deviation of pressure (indicated by background shading) is somewhat increased when dimensions are low, which may well correspond to the previous observations concerning the amplitude of the leading wavenumber-two pattern (figure \ref{fig:EVs_and_cor}\,a and b).\par
Somewhat surprisingly, the corresponding mean state in the upper tail of the $\locdim$-distribution is nearly identical to the climatology (cf. figure \ref{fig:PC_extreme}\,a and \ref{fig:PC_extreme_geo}\,c). A very slight pressure increase over the Northern and decrease over the Southern Atlantic constitute the only notable anomaly. The zonal asymmetry represented by the third principal component, which clearly dominates low dimensional states, makes no significant contribution at all.
\begin{figure}[!t]
	\centering
	\includegraphics[width=\textwidth]{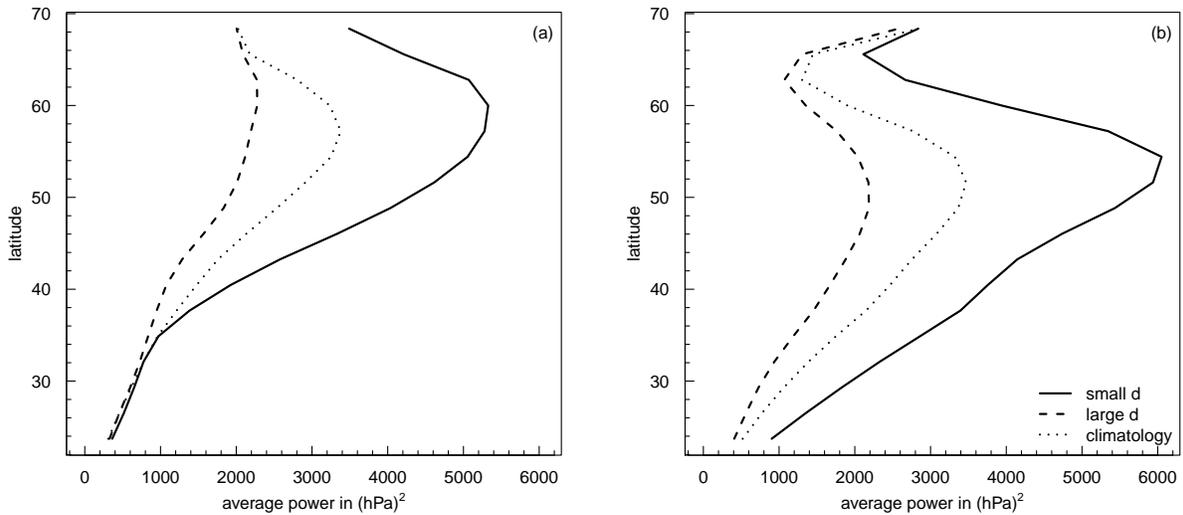}
	\caption{Average zonal power spectrum for wave number one (a) and two (b) as a function of latitude, given that $\locdim$ is in the upper 2\,\% (dashed) or lower 2\,\% (solid) of its distribution. Climatological mean spectra shown as dotted lines.} 
	\label{fig:fft}
\end{figure}
We can obtain a clearer view of this phenomenon by considering zonal power spectra: For each latitude and time, we take the squared absolute value of the field's Fourier transform as a measure of the energy on different spatial scales. These spectra, averaged over states with extreme local dimension (as in figure \ref{fig:PC_extreme_geo}) are partly displayed in figure \ref{fig:fft}. When $\locdim$ is large, the power contained in patterns of wavenumber one (meridional dipole) and two (Rossby waves) is systematically decreased compared to the climatological mean. The reverse is true for extremely small dimensions, which are associated with enhanced wave-motion (compare the standard deviations in figure \ref{fig:PC_extreme_geo}\,b). The anticorrelation between $\locdim$ and the contribution of large waves is of course consistent with our observations concerning the leading principal components (figure \ref{fig:PCcors}), as well as the roughly triangular shape of the bivariate histograms seen in figure \ref{fig:EVs_and_cor}. 

\section{Discussion}\label{sec:dis}

This study has explored some aspects of the newly available access to local attractor dimensions via the statistics of extreme recurrences. We have demonstrated that the numerical estimates of previous authors can, in principle, be replaced by an analytical maximum likelihood estimator of the local dimension $\locdim$. This step results in a simplified procedure, removes the somewhat arbitrary choice of a distance observable and allows for an explicit expression of asymptotic uncertainty. It furthermore enables an additional test of agreement between theoretically expected and experimentally observed behavior, at the cost of re-introducing the choice of observable. 
As an atmosphere-like test case, we have employed the PUMA model, using daily output over one millennium. With its absence of randomness and transience (due to a deactivated seasonal cycle), this system represents a suitable intermediate step between simple low-dimensional attractors and the truly realistic representations of the atmosphere investigated by previous authors. Our analysis was deliberately limited to the same truncation of state vectors used by Ref.\,\onlinecite{faranda2017} (North Atlantic sea level pressure), thus enabling comparisons.\par  
We find that the new estimator yields systematically increased values, which might be interpreted as an improvement: All our estimates of average attractor dimensions rise as we move to more extreme recurrences while the statistical tests simultaneously indicate increasing agreement with theory. These observations suggest that the mean local dimension very slowly approaches a true limit value on the order of $D_1\approx 10^2$, as found by Ref.\,\onlinecite{cruz2018}. In this interpretation, the novel estimator has a slightly reduced negative bias. The results of Refs.\,\onlinecite{cruz2018} and \onlinecite{galfi2017} however indicate that the true global dimension is far away and would only be attainable from significantly longer trajectories (on the order of millions of years), far exceeding the computational resources available to us.\par 
Beyond estimates of the global information dimension, we are mostly interested in the possibility of ranking individual system states according to their observed local dimension, thus quantifying the recurrence behavior of various points on the attractor. We have found that this ordering is largely invariant to the choice of estimator, since analytically and numerically derived time-series of $\locdim$ remain highly correlated in all tested situations. It can thus be argued that the new procedure is preferable in most applications due to its desirable properties listed above. Note, however, that this method can suffer from considerable errors if the time-series of distances is short and contains several direct predecessors and successors of $x_i$. These non-recurrences can dominate our estimate if the number of truly extreme returns to $x_i$ is sufficiently low (cf. figure \ref{fig:Dpoftau_30d}) and should therefore be discarded as a cautionary measure.\par
In a second step, we have investigated the connection between local dimensions and patterns of circulation. 
Using the information on $\locdim$ in conjunction with classical techniques like PCA and Fourier analysis, we were able to identify an Atlantic dipole structure, which is associated with low local dimensions and thus more frequent recurrences. It is interesting to note that the presence of such a pattern is not evident in the climatological mean field and could not be inferred from the EOFs alone, since the shape of these vectors is artificially constrained by orthogonality. 
On the opposite end of $\locdim$'s distribution, we found states with overall weak pressure anomalies. More specifically, large local dimensions correlate with a decrease in Rossby wave activity. 
These results are in good agreement with our intuition about local attractor dimensions: When a few large waves dominate the system's behavior, its individual states can be described by only a handful of numbers, namely the phases and amplitudes of those waves. In the absence of such large, dominant patterns, the system can evolve in numerous directions, corresponding to high local dimensions. It is nonetheless somewhat surprising that no obvious link between high local dimensions and blocked flow-configurations was found. Atmospheric blocking patterns are known to occur in the model\cite{cruz2018} and have previously been shown to correspond to high instability\cite{faranda2017,schubert2016}. Their missing connection to high values of $\locdim$ in the present investigation remains unexplained for the time being.\par
Concluding this study, we note that our observations are broadly consistent with the results for reanalysis data sets presented by Ref.\,\onlinecite{faranda2017}, Ref.\,\onlinecite{messori2017} and others. Disagreement about the specific patterns associated with high- and low-dimensional states is not unexpected because the PUMA model's ability to reproduce realistic atmospheric motion is fundamentally limited -- particularly since only one phase of the seasonal cycle was considered. We have however demonstrated that the link between $\locdim$ and particular circulation regimes can be observed in the absence of periodic or stochastic forcing and is not a statistical artifact. This lends further credence to the claim that estimates of $\locdim$ characterize domains of atmospheric motion and may be of use in identifying such patterns. How these observations might change with longer integrations as the mean local dimension approaches the true limit value of $D_1$, is an interesting question for future research.\par

\begin{acknowledgments}
	We would like to thank Andreas Hense for several helpful discussions. We are furthermore grateful to two anonymous reviewers for their constructive criticism and valuable suggestions.
\end{acknowledgments}

\appendix

\section{Connection between local and continuous intrinsic dimension}\label{app:intdim}
Ref.\,\onlinecite{houle2015} defines the \textit{continuous intrinsic dimension} as follows:
\begin{align}
\intdim_i(r)=\lim\limits_{\alpha\to 0} \frac{\ln F_i(r(1+\alpha))-\ln F_i(r)}{\ln(1+\alpha)}\comgood\label{eq:IDx_def}
\end{align}
where $F_i$ is the unconditional cdf of the distances $r$ (interpreted here as \textit{search radii}) to some point $x_i$. Assuming differentiability of $F_i$, it can be shown that $\intdim_i$ is in fact equivalent to a particular definition of indiscriminability, i.e. the distance measure's inability to differentiate between two similar states. The authors furthermore find that, in the limit of two small distances $\epsilon_1$ and $\epsilon_2$, the cdf satisfies
\begin{align}
F_i(\epsilon_1)/F_i(\epsilon_2)\to (\epsilon_1/\epsilon_2)^{\intdim_i(0)}\dotgood\hspace{10pt}\text{(Theo. 6, 7 in Ref.\,\onlinecite{houle2015})}\label{eq:IDx}
\end{align}
From this, it is easy to see that $\intdim_i$ converges to the local dimension $\locdim(x_i)$ (equation \ref{eq:Dp}) in the limit of small distances: Setting $\epsilon_2$ to some small constant value and recalling that $F_i(r)=\mu(B_r(x_i))$, we realize that equations \ref{eq:Dp} and \ref{eq:IDx} in fact describe the exact same scaling law, implying $\intdim_i(0)=\locdim(x_i)$. This result can also be obtained directly by introducing $\locdim$ into equation \ref{eq:IDx_def} after taking the limit $r\to 0$.\par

%

\end{document}